\documentclass[aps,prc,showpacs,twocolumn,superscriptaddress]{revtex4-1}

\usepackage{graphicx}
\usepackage{longtable}

\begin{document}

\title{Low lying states in $^{8}$B}

\author{J.P. Mitchell}
\email{jpmitchell@fsu.edu}
\affiliation{Department of Physics, Florida State University, Tallahassee, Florida 32306, USA}

\author{G.V. Rogachev} 
\email{grogache@fsu.edu}
\affiliation{Department of Physics, Florida State University, Tallahassee, Florida 32306, USA}

\author{E.D. Johnson}
\affiliation{Department of Physics, Florida State University, Tallahassee, Florida 32306, USA}

\author{L.T. Baby}
\affiliation{Department of Physics, Florida State University, Tallahassee, Florida 32306, USA}

\author{K.W. Kemper}
\affiliation{Department of Physics, Florida State University, Tallahassee, Florida 32306, USA}

\author{A.M. Moro}
\affiliation{Departamento de FAMN, Facultad de F'sica, Universidad de Sevilla, Apartado 1065, E-41080 Sevilla, Spain}

\author{P.N. Peplowski}
\altaffiliation{Present address: Applied Physics Laboratory, Johns Hopkins University, Laurel, 20723 MD}
\affiliation{Department of Physics, Florida State University, Tallahassee, Florida 32306, USA}

\author{A. Volya}
\affiliation{Department of Physics, Florida State University, Tallahassee, Florida 32306, USA}

\author{I. Wiedenh\"over}
\affiliation{Department of Physics, Florida State University, Tallahassee, Florida 32306, USA}

\date{\today}

\begin{abstract}
Excitation functions of elastic and inelastic $^{7}$Be+p scattering were measured in the energy range between 1.6 and 2.8 MeV in the c.m. An R-matrix analysis of the excitation functions provides strong evidence for new positive parity states in $^{8}$B. A new 2$^+$ state at an excitation energy of 2.55 MeV was observed and a new 0$^+$ state at 1.9 MeV is tentatively suggested. The R-matrix and Time Dependent Continuum Shell Model were used in the analysis of the excitation functions. The new results are compared to the calculations of contemporary theoretical models.
\end{abstract}


\maketitle

Light nuclei are of great importance in modern nuclear physics as their structure provides a link between nucleon-nucleon interactions and macroscopic nuclear many-body dynamics. Generally, properties of stable light nuclei, including level schemes, are reproduced rather well by the so-called \textit{ab initio} methods that start from the basic interactions of nucleons  \cite{Navratil09,Pieper04}. However, the neutron deficient isotope of boron, $^{8}$B, and its mirror nucleus $^{8}$Li, provide an interesting exception. Most \textit{ab initio} calculations, including one from more than 10 years ago \cite{Navratil98}, predict more positive parity states below 4 MeV than what has been observed experimentally. These ``missing'' states are proposed to have a relatively simple structure, with large spacing between the levels and at low excitation energies. Consequently, it is rather surprising that these levels have not been observed to date. The importance of the $^{7}$Be(p,$\gamma$)$^{8}$B reaction for understanding the solar neutrino flux is another stimulus for taking a closer look at the $^{8}$B structure. Previously unaccounted for low-lying states in $^{8}$B may alter the theoretical extrapolation of the $^{7}$Be(p,$\gamma$)$^{8}$B S-factor. The main objective of this work is an experimental search for these proposed low-lying levels in $^{8}$B.

\begin{figure}
   \includegraphics[width=1.0\linewidth]{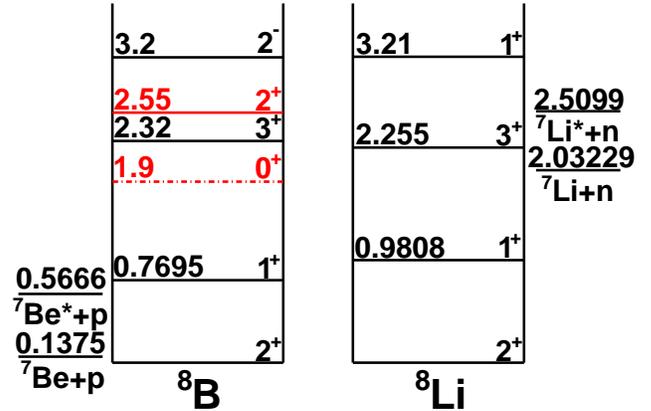}
   \caption{ \label{fig:level_scheme}(Color online) Experimental level scheme of $^{8}$B and $^8$Li at excitation energies below 4 MeV. New states are shown in red. The dash-dotted line indicates that the state is tentative.}
\end{figure}

The level structure of $^8$B and $^8$Li below 4 MeV is shown in Figure \ref{fig:level_scheme}. The first and second excited states, the 1$^{+}$ at 0.7695 MeV and the 3$^{+}$ at 2.32 MeV, have been observed in numerous experiments, reviewed in \cite{Tilley04}. The broad negative parity state at $\sim$3.0 MeV was first suggested in Ref. \cite{Goldberg98} and later identified as a 2$^{-}$ state at 3.5$\pm$0.5 MeV in Ref. \cite{Rogachev01}. An excitation function of $^{7}$Be+p resonance elastic scattering was measured in both of these works. The most recent $^{7}$Be+p measurement, performed by H. Yamaguchi, et al., \cite{Yamaguchi09}, confirmed the 2$^{-}$ state and determined the excitation energy and width of this state with better precision, E=3.2$_{-0.2}^{+0.3}$ MeV and $\Gamma$=3.4$_{-0.5}^{+0.8}$ MeV. No new positive parity states in the $^{7}$Be+p elastic scattering excitation function below 4 MeV have been observed in any of these studies. However, it is possible that the missing states still exist but contribute very little to the elastic excitation function due to the strong decay branch to the 1/2$^{-}$ first excited state of $^{7}$Be at 0.43 MeV. This possibility was considered by D. Halderson in the framework of the Recoil Corrected Continuum Shell Model (RCCSM) \cite{Halderson04}, and the suggestion was made to use inelastic $^{7}$Be+p scattering to search for the missing states. It is interesting to point out that  some evidence for the 2$^+$ state at 3.0 MeV was also presented in \cite{Halderson04} based on the RCCSM analysis of the $^{7}$Be+p elastic scattering excitation function measured in \cite{Rogachev01}. In the present work, the excitation functions of the $^{7}$Be+p elastic and inelastic scattering were measured simultaneously and a consistent R-matrix analysis of both excitation functions was performed.

The nucleus $^8$B is weakly bound with a proton separation energy of only 137 keV. All of the excited states of this nucleus are in the continuum. The recently developed Time Dependent Continuum Shell Model (TDCSM) approach \cite{Volya2009} bridges the reaction-structure gap. Within this approach the cross sections for elastic and inelastic nucleon scattering can be calculated directly from the nuclear effective Hamiltonian. Influence of the continuum on the wavefunctions of the populated resonances is treated self-consistently and the number of free parameters is greatly reduced. Once residual interactions are chosen, only one free parameter (excitation energy) remains for each resonance (compared to five in two-channel R-matrix approach). TDCSM analysis of the resonance scattering data is presented here. It was used not only as a stand-alone tool but also as a logical starting point for the subsequent R-matrix fit.

\begin{figure}
\includegraphics[width=.9\linewidth]{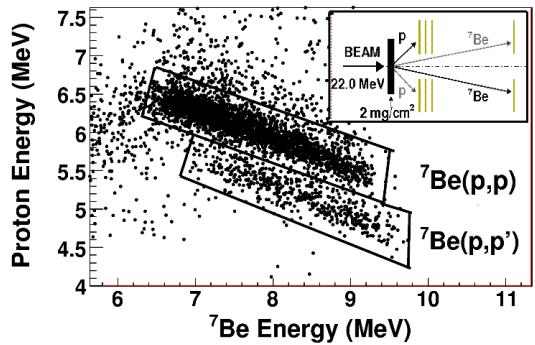}
\caption{\label{fig:2d} (Color online) Scatter plot of kinematic coincidence between detected protons and $^7$Be ions. Regions which correspond to elastic and inelastic scattering are labeled. The schematic view of the experimental setup with a sandwich of three Micron Semiconductor S2 detectors for the light recoils and a single S2 detector for the heavy recoils is shown in the inset.} 
\end{figure}

The experiment was carried out at the John D. Fox Superconducting Accelerator Laboratory at Florida State University. A radioactive beam of $^{7}$Be was produced using the $^{1}$H($^{7}$Li,$^{7}$Be)n reaction. The primary $^{7}$Li beam was accelerated by a 9MV SuperFN Tandem Van de Graaff accelerator followed by a LINAC booster. The primary target was a 4 cm long hydrogen gas cell with 2.5 $\mu$m Havar entrance and exit windows. The gas cell was cooled by liquid nitrogen and had a gas pressure of 390 mBar. The radioactive nuclear beam facility RESOLUT was used to separate $^{7}$Be from other reaction products and the primary beam. Two $^{7}$Be beam energies were used in this experiment: 22.0, and 18.5 MeV. The typical intensity of the $^{7}$Be beam was $10^{5}$ pps. The composition of the beam was 70$\%$ $^{7}$Be with 30$\%$ $^{7}$Li contaminant.

A sketch of the experimental setup is shown in the inset of Figure \ref{fig:2d}. A solid polyethylene (C$_{2}$H$_{4}$) target of thickness optimized for the given beam energy (see description below) was used. A set of three annular Micron Semiconductor silicon strip detectors (S2 design) for the proton recoils were positioned 5, 6 and 7 cm downstream from the target. Another S2 detector for the $^{7}$Be recoils was positioned 24.5 cm from the target. The S2 detector has annular geometry and consists of 16 segments and a side of rings that allow for the scattering angle of the products to be determined. The first in the set of three proton detectors was a $\Delta$E detector with thickness 65 $\mu$m. It was only used in the initial part of the experiment to verify that correct identification of light recoils can be achieved. The other two proton detectors and the $^{7}$Be detector were 500 $\mu$m each in thickness.

The target thickness was optimized for maximum energy losses of the $^{7}$Be ions in the target while ensuring that all $^{7}$Be recoils make it out of the target with enough kinetic energy left to be detected in the downstream S2 detector. Kinematic coincidence between protons in the array of three S2 detectors and the $^{7}$Be recoils in the downstream S2 detector was then used to identify the scattering events. The time between the events in the proton and $^{7}$Be detectors was measured with time resolution of about 3 ns to eliminate random coincidence background. Elastic and inelastic scattering processes can be distinguished because the complete kinematics of the event is measured. More detailed description of the experimental technique can be found in \cite{Rogachev2010}.

The 2D scatter plot for the kinematic coincidence between protons and $^{7}$Be is shown in Figure \ref{fig:2d}. The kinematic loci which correspond to elastic and inelastic scattering processes are labeled and outlined with contours. Polyethylene target thicknesses used in this experiment were 2.5 and 1.5 mg/cm$^{2}$ for the 22 and 18.5 MeV beam energies respectively. In addition, a separate run at 18.5 MeV of $^{7}$Be beam energy was performed with a slightly thicker (2 mg/cm$^{2}$) target to extend the measured excitation function to lower energies without changing the energy of the beam. This use of a thicker target comes at the price of losing coincidences between the highest energy protons and the $^{7}$Be recoils because they are produced at the very beginning of the target and never make it through. Only the lower energy part of this spectrum was used in the analysis.

\begin{figure}
\includegraphics[width=1\columnwidth]{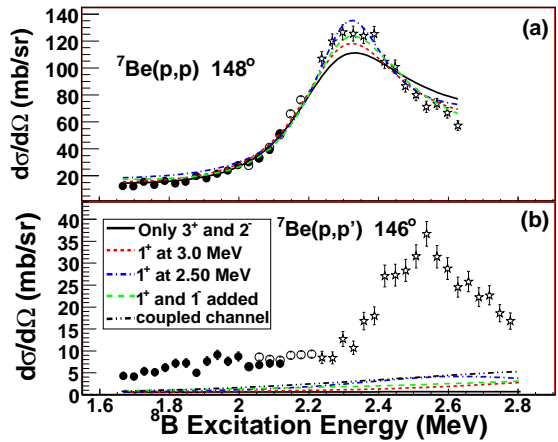} 
\caption{\label{fig:nonew} (Color online) Excitation functions of $^7$Be+p elastic (top) and inelastic (bottom) scattering. Data from three different runs are shown. The open and solid circles correspond to the run at 18.5 MeV of $^{7}$Be with a 1.5 mg/cm$^{2}$ and 2 mg/cm$^{2}$ targets respectively, and the stars are at 22 MeV with a 2.5 mg/cm$^{2}$ target. The black solid curve is an R-matrix fit with only the known 3$^+$ and 2$^-$ states at 2.3 and 3.5 MeV respectively. The red short-dashed line includes contribution of the higher lying 1$^+$ states assumed at 3.0 MeV. The blue dash-dotted line shows the 1$^+$ state shifted to 2.5 MeV and the green long-dashed line also includes the 1$^-$ state introduced at 5 MeV. The black dash-double-dotted curve in the bottom panel shows the excitation function of $^7$Be(p,p$'$)$^7$Be(0.43) reaction from the direct mechanism.}
\end{figure}

Figure \ref{fig:nonew} shows excitation functions of elastic (top panel) and inelastic (bottom panel) scattering of $^{7}$Be+p measured in three different runs. The open circles correspond to the run at 18.5 MeV of $^{7}$Be with a 1.5 mg/cm$^{2}$ target, the solid circles are 18.5 MeV of $^{7}$Be with a 2 mg/cm$^{2}$ target data and the stars are 22 MeV of $^{7}$Be with 2.5 mg/cm$^{2}$ target data. Binning of 4$^{\circ}$ in the lab frame was used. Absolute normalization of the cross section was done using the known excitation functions for the $^{7}$Li+p elastic scattering \cite{Walters1953}. These excitation functions were extracted from the experimental data using the same procedure as for the $^{7}$Be+p elastic scattering. Excitation functions extracted from our data agree well with the differential cross section for the elastic and inelastic scattering of $^{7}$Be+p measured at several energies of $^{7}$Be using the conventional thin target approach and reported by U. Greife, et al., \cite{Greife2007}.

A striking feature of our data is that the cross section for inelastic scattering is very large ($\sim$30 mb/sr at an excitation energy of 2.5 MeV). Two channel, multi-level R-matrix analysis clearly indicates that it is not possible to explain this high inelastic cross section if only known states in $^{8}$B are considered (Figure \ref{fig:nonew}). This failure can be understood from the following simple considerations. The first excited 1$^{+}$ state at 0.77 MeV is too narrow (35.6 keV) to have any significant impact on the excitation functions at energies above 1.5 MeV. The second excited state, 3$^{+}$ at 2.32 MeV, can only decay to the $1/2^{-}$ first excited state of $^{7}$Be with orbital angular momentum $\ell=3$. Therefore, even if the corresponding reduced width is large, the inelastic partial proton width, $\Gamma_{p'}=2P_{\ell}(kR)\gamma^{2}$, would still be small compared to the elastic partial proton width due to a small penetration factor for high angular momentum decay. Hence, the cross section for population of the first excited state in $^{7}$Be due to the 3$^{+}$ resonance in $^{8}$B, determined by the $\Gamma_{p}\Gamma_{p'}/\Gamma_{\rm tot}^{2}$ ratio, is small. The same is true for the broad 2$^{-}$ state in $^{8}$B at 3.2 MeV as it can only decay to the first excited state in $^{7}$Be with angular momentum $\ell=2$ while decay to the g.s. proceeds with $\ell=0$. The black solid curve in Figure \ref{fig:nonew} shows the results of an R-matrix calculation with only the previously known 1$^{+}$, 3$^{+}$ and 2$^{-}$ states at 0.77, 2.32 and 3.7 MeV with reduced widths parameters evaluated using the TDCSM approach and known total widths of these states. (Excitation energy and width of the 2$^-$ were adjusted slightly to produce a better fit.) It is clear that while the elastic scattering data are well reproduced, the inelastic scattering data cannot be explained by the known states.

An attempt has been made to reproduce the observed p+$^7$Be inelastic scattering excitation function without introducing new resonances in $^8$B but assuming a direct excitation of the $^7$Be first excited state in p+$^7$Be scattering. In this case the reaction does not proceed through the population of resonances in $^8$B and cannot be evaluated using R-matrix approach. The calculations were performed using the  coupled-channels approach with the computer code {\sc fresco} \cite{fresco}. The black dash-double-dotted curve in the bottom panel of Figure \ref{fig:nonew} shows the excitation function of the $^7$Be(p,p$'$)$^7$Be$^*$(0.43) inelastic scattering at 146$^{\circ}$ due to the direct mechanism. It is clear that unless additional resonance(s) are introduced direct excitation cannot be responsible for the large inelastic scattering cross section observed experimentally. Details of this calculation will be published elsewhere.

Based on the level scheme of $^{8}$Li it is natural to try to introduce the second 1$^{+}$ state in $^{8}$B at an excitation energy around 3 MeV. Reduced widths for this state were chosen according to the TDCSM calculations carried out with Cohen-Kurath interaction (CKI) \cite{Cohen1965}. It was verified that these reduced widths reproduce the known width of this state in $^8$Li ($\sim$1 MeV). The TDCSM predicts that this state has a substantial inelastic partial width. The red short-dashed curve in Figure \ref{fig:nonew} shows the effect of the 1$^+$ state at 3.4 MeV on the fit. While the elastic excitation function is fitted perfectly, the inelastic cross section is still underestimated, even if this state is shifted to 2.5 MeV, where inelastic scattering has its maximum cross section (blue dash-dotted curve in Figure \ref{fig:nonew}). Finally, in an attempt to increase the inelastic cross section without using new states below 3 MeV we introduced a ``background'' state, the 1$^-$ at $\sim$5 MeV. This state is a spin-orbit partner to the known 2$^-$ state and splitting between these states should be 1-2 MeV based on TDCSM calculations.   
This state can decay to the first excited state of $^7$Be with $\ell=0$, therefore it may contribute significantly to the inelastic cross section. The reduced widths for the 1$^-$ state were evaluated using the TDCSM (WBP \cite{Warburton1992} residual interactions were used). As expected, the 1$^-$ state increases the inelastic cross section overall (green long-dashed curve in Figure \ref{fig:nonew}). But even with this state included the inelastic cross section cannot be reproduced.

\begin{figure*}
\includegraphics[width=0.9\textwidth]{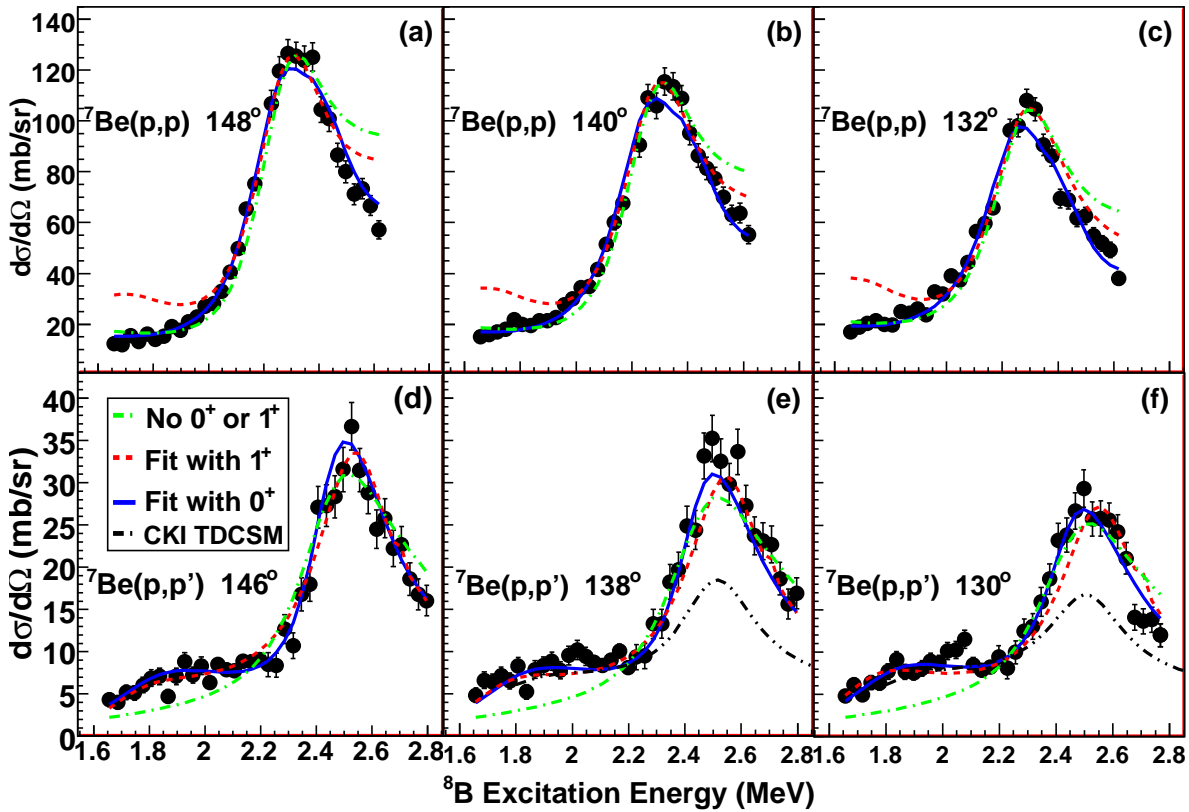} 
\caption{\label{fig:final}(Color online) Final R-matrix fit of the experimental data, with the new
2$^{+}$ and 0$^{+}$ states. $^7$Be+p elastic scattering data at three c.m. angles are shown in the top panels (a-c) and inelastic data are shown in the bottom panels (d-f). Solid blue line is the best fit, dashed red line is a fit with 1$^+$ at 1.9 MeV instead of 0$^+$  and green dash-dotted line is a fit without the 0$^+$. Dashed-double-dotted cyan line in the bottom panel is TDCSM calculations with the known states in $^8$B and the new 0$^+$ and 2$^+$ at 1.9 and 2.55 MeV respectively.}
\end{figure*}

\begin{table}
\caption{\label{tab:states} Parameters of resonances in $^8$B from the R-matrix best fit. States in parenthesis are unknown states outside of the measured excitation energy range. They provide essential ``background'' through low energy tails. New states in $^8$B introduced in this work indicated with superscript (a). 
Uncertainties correspond to 1$\sigma$.}
\begin{ruledtabular}
\begin{tabular}{ccccc}
J$^{\pi}$ & E$_{ex}$ & $\Gamma_{tot}$ & $\Gamma_{p}$ & $\Gamma_{p'}$ \\
\hline
$1^{+}$ & 0.7695 & 0.035 & 0.034 & 0.001 \\
$^{a}0^{+}$ & 1.9(1) & 0.61(15) & 0.28(14) & 0.33(18) \\
$3^{+}$ & 2.28(2) & 0.34(3) & 0.34(3) & 0.0 \\
$^{a}2^{+}$ & 2.55(2) & 0.36(12) & 0.12(4) & 0.24(11) \\
$^{a}(1^{+})$ & 3.4 & 1.34 & 1.16 & 0.18 \\
$2^{-}$ & 3.8 & 4.7 & 4.7 & 0.0 \\
$^{a}(1^{-})$ & 5.1 & 4.6 & 2.3 & 2.3 \\
\end{tabular}
$^{a}$ New levels suggested in this work. 
\end{ruledtabular}
\end{table}

The ``extra'' low-lying states in $^8$B predicted by {\it ab initio} and shell model calculations have spin-parity assignments 0$^+$, 1$^+$ and 2$^+$. Influence of the 1$^+$ state has already been discussed. It was found that introducing a new 2$^{+}$ state placed at 2.55 MeV, reproduces both the magnitude and angular dependence of the observed peak in the inelastic cross section while keeping the elastic excitation function in agreement with the experimental data. This is shown as green dash-dotted curve in Figure \ref{fig:final}, where $^7$Be+p elastic (panels a-c) and inelastic (panels d-f) scattering excitation functions measured at three c.m. angles are shown. However, even with this new 2$^+$ state the cross section for inelastic scattering below 2.3 MeV is still too low. The 2$^{+}$ state should have a relatively small width (360 keV) to fit the observed peak-like structure in the inelastic excitation function at 2.5 MeV, and its influence below 2.3 MeV is small. The only predicted state which has not been considered is the 0$^{+}$. Introducing a new 0$^{+}$ state at the excitation energy of 1.9 MeV with a width of 610 keV allows the inelastic scattering data to be fit below 2.3 MeV without destroying the fit to the elastic scattering data (solid line in Figure \ref{fig:final}). It was verified that a 1$^{+}$ spin-parity assignment for this new state is not possible as it will ruin agreement with the elastic scattering data (red dashed line in Figure \ref{fig:final}). An important distinction between the 0$^+$ and the 2$^+$ states has to be made. While existence of the 2$^+$ state is hard to dismiss, the case for the 0$^+$ state is somewhat weaker and further investigation is warranted. 

An important role in guiding our analysis and providing support for the 0$^+$ state was played by the TDCSM calculations \cite{Volya2009}. The excitation function for $^7$Be(p,p$'$)$^7$Be(0.43) inelastic scattering calculated using the TDCSM approach is shown in the bottom panels of Figure \ref{fig:final} as a dash-double-dotted cyan curve. The only free parameters in these calculations are the excitation energies of the states. All known states in $^8$B and the new 0$^+$ and 2$^+$ states at 1.9 and 2.55 MeV were taken into account in these calculations. The CKI residual interaction \cite{Cohen1965} was used and the corresponding states were shifted to their experimental locations. Note that the cross section at $\sim$2 MeV is well reproduced by the 0$^+$ state. This can be considered as an additional argument in favor of the 0$^+$ state at 1.9 MeV in $^8$B.  At $\sim$2.5 MeV the TDCSM cross section is determined by the 2$^+$ state and it is lower than observed experimentally. This is due partially to the absence of the negative parity states (specifically the 1$^-$ state)   in the applied TDCSM model space. Details of the TDCSM analysis will be published elsewhere.

\begin{figure}[t]
\includegraphics[width=0.8\columnwidth]{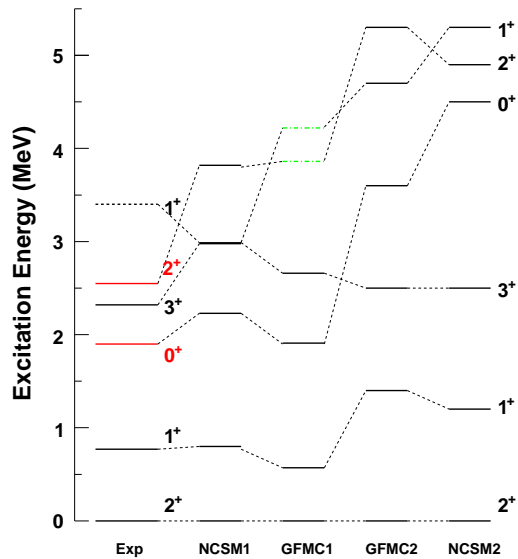} 
\caption{\label{fig:abinitio}(Color online) Comparison of the experimental data to the predictions of the {\it ab initio} models. The subset ``Exp'' shows experimental data. The new states in $^8$B observed in this experiment are shown in red. The dashed line is a 1$^+$ state inferred from level structure of $^8$Li based on mirror symmetry. The subset ``NCSM1'' is the result of the most recent NCSM calculations with the CD-Bonn 2000 potential \cite{Navratil2006}. The subset ``GFMC1'' is the earlier GFMC result \cite{wiri00} with AV18/UIX Hamiltonians (values are for $^8$Li and excitation energies for the last two states, shown as green dash-dotted lines, were calculated using Variational Monte Carlo rather than GFMC method). The subset ``GFMC2'' is the more recent GFMC prediction \cite{Pieper04} with AV18/IL2 Hamiltonians (values are for $^8$Li). The subset ``NCSM2'' is the earlier NCSM result \cite{Navratil98}.}
\end{figure}

Table \ref{tab:states} shows the best fit parameters of the 7 states, which were introduced to describe both elastic and inelastic data simultaneously. Three of these states, the first 1$^+$, 3$^+$ and 2$^-$ were known. The two states shown in parenthesis are unknown ``background'' states. The initial fit parameters for these states were taken from the TDCSM calculations and also (in the case of the 1$^+$ at 3.4 MeV) from the level structure of the mirror nucleus, $^8$Li. The new low-lying 0$^{+}$ and 2$^{+}$ states were introduced to fit the measured cross section for p+$^7$Be inelastic scattering. The location of these states in $^8$Li is not known, calling for a new experimental effort to locate them in $^8$Li.

Comparison of the experimental spectrum of $^8$B with the predictions of {\it ab initio} models is shown in Figure \ref{fig:abinitio}. Clearly, there is no unified picture for the level structure of $^8$B - $^8$Li isotopes from the available array of {\it ab initio} calculations. There is a certain general trend, however. All of them produce a 0$^+$ as the second or third excited state, always below the known 1$^+_{2}$ state (experimentally found at 3.2 MeV in $^8$Li) and the 2$^+$ state, found in this work at 2.55 MeV. Our experimental result seems to confirm this prediction. The 2$^+$ state is generally found at higher excitation energy in {\it ab initio} calculations than observed in this work for $^8$B.

In summary, the excitation function for p+$^7$Be elastic and inelastic scattering was measured in the energy range of 1.6 - 2.8 MeV and the c.m. angular range of 132 - 148 degrees. The R-matrix analysis of the excitation functions indicates that new low-lying states in $^8$B have to be introduced to explain the large inelastic cross section with a well defined peak at 2.5 MeV. These new states are suggested to be the 0$^+$ at 1.9 MeV and 2$^+$ at 2.55 MeV with width 610 keV and 340 keV respectively. Evidence for the 2$^+$ state at 2.55 MeV is reliable. The 0$^+$ at 1.9 MeV can be considered as tentative due to uncertainties associated with the coupled-channels calculations and possible contributions from the tails of the higher energy resonances into the inelastic scattering cross section. Accurate measurement of the p+$^7$Be excitation function for inelastic scattering in the energy range from 0.7 to 2.0 MeV and in a broad angular range should provide a definitive answer on the existence of the 0$^+$ resonance.

The TDCSM analysis of the $^7$Be+p scattering has been performed. The TDCSM reduces the number of free parameters in the fit and links directly nuclear structure to nuclear reaction cross sections while treating the continuum self-consistently. The role of TDCSM in our work was to provide important guidelines for constraining the R-matrix fit. We believe that it is an essential tool for analysis of resonance scattering data.

Comparing new experimental results to the predictions of the {\it ab initio} models \cite{Navratil98,wiri00,Pieper04,Navratil2006} we notice that  there is no unified picture for the level structure of $^8$B - $^8$Li isotopes. We hope that new experimental data on the structure of exotic nuclei (including those presented here) will serve as a guide for construction of more accurate {\it ab initio} models.

The authors are grateful to John Schiffer, Donald Robson and John Millener for criticism and enlightening discussions and to Vladilen Goldberg for continuously stimulating interest in this work and for reading the manuscript and making important corrections and suggestions. The work was supported by the National Science Foundation under grant number PHY-456463, by the U.S. Department of Energy grant DE-FG02-92ER40750, and the state of Florida.

\bibliography{8B}

\begin{thebibliography}{17}%
\makeatletter
\providecommand \@ifxundefined [1]{%
 \@ifx{#1\undefined}
}%
\providecommand \@ifnum [1]{%
 \ifnum #1\expandafter \@firstoftwo
 \else \expandafter \@secondoftwo
 \fi
}%
\providecommand \@ifx [1]{%
 \ifx #1\expandafter \@firstoftwo
 \else \expandafter \@secondoftwo
 \fi
}%
\providecommand \natexlab [1]{#1}%
\providecommand \enquote  [1]{``#1''}%
\providecommand \bibnamefont  [1]{#1}%
\providecommand \bibfnamefont [1]{#1}%
\providecommand \citenamefont [1]{#1}%
\providecommand \href@noop [0]{\@secondoftwo}%
\providecommand \href [0]{\begingroup \@sanitize@url \@href}%
\providecommand \@href[1]{\@@startlink{#1}\@@href}%
\providecommand \@@href[1]{\endgroup#1\@@endlink}%
\providecommand \@sanitize@url [0]{\catcode `\\12\catcode `\$12\catcode
  `\&12\catcode `\#12\catcode `\^12\catcode `\_12\catcode `\%12\relax}%
\providecommand \@@startlink[1]{}%
\providecommand \@@endlink[0]{}%
\providecommand \url  [0]{\begingroup\@sanitize@url \@url }%
\providecommand \@url [1]{\endgroup\@href {#1}{\urlprefix }}%
\providecommand \urlprefix  [0]{URL }%
\providecommand \Eprint [0]{\href }%
\@ifxundefined \urlstyle {%
  \providecommand \doi  [0]{\begingroup \@sanitize@url \@doi}%
  \providecommand \@doi [1]{\endgroup \@@startlink {\doibase
  #1}doi:\discretionary {}{}{}#1\@@endlink }%
}{%
  \providecommand \doi  [0]{doi:\discretionary{}{}{}\begingroup
  \urlstyle{rm}\Url }%
}%
\providecommand \doibase [0]{http://dx.doi.org/}%
\providecommand \Doi [0]{\begingroup \@sanitize@url \@Doi }%
\providecommand \@Doi  [1]{\endgroup\@@startlink{\doibase#1}\@@Doi}%
\providecommand \@@Doi [1]{#1\@@endlink}%
\providecommand \selectlanguage [0]{\@gobble}%
\providecommand \bibinfo  [0]{\@secondoftwo}%
\providecommand \bibfield  [0]{\@secondoftwo}%
\providecommand \translation [1]{[#1]}%
\providecommand \BibitemOpen [0]{}%
\providecommand \bibitemStop [0]{}%
\providecommand \bibitemNoStop [0]{.\EOS\space}%
\providecommand \EOS [0]{\spacefactor3000\relax}%
\providecommand \BibitemShut  [1]{\csname bibitem#1\endcsname}%
\bibitem [{\citenamefont {Navratil}\ \emph {et~al.}(2009)\citenamefont
  {Navratil}, \citenamefont {Quaglioni}, \citenamefont {Stetcu},\ and\
  \citenamefont {Barrett}}]{Navratil09}%
  \BibitemOpen
  \bibfield  {author} {\bibinfo {author} {\bibfnamefont {P.}~\bibnamefont
  {Navratil}}, \bibinfo {author} {\bibfnamefont {S.}~\bibnamefont {Quaglioni}},
  \bibinfo {author} {\bibfnamefont {I.}~\bibnamefont {Stetcu}}, \ and\ \bibinfo
  {author} {\bibfnamefont {B.~R.}\ \bibnamefont {Barrett}},\ }\href@noop {}
  {\bibfield  {journal} {\bibinfo  {journal} {J. Phys. G},\ }\textbf {\bibinfo
  {volume} {36}},\ \bibinfo {pages} {083101} (\bibinfo {year}
  {2009})}\BibitemShut {NoStop}%
\bibitem [{\citenamefont {Pieper}\ \emph {et~al.}(2004)\citenamefont {Pieper},
  \citenamefont {Wiringa},\ and\ \citenamefont {Carlson}}]{Pieper04}%
  \BibitemOpen
  \bibfield  {author} {\bibinfo {author} {\bibfnamefont {S.~C.}\ \bibnamefont
  {Pieper}}, \bibinfo {author} {\bibfnamefont {R.~B.}\ \bibnamefont {Wiringa}},
  \ and\ \bibinfo {author} {\bibfnamefont {J.}~\bibnamefont {Carlson}},\
  }\href@noop {} {\bibfield  {journal} {\bibinfo  {journal} {Phys. Rev. C},\
  }\textbf {\bibinfo {volume} {70}},\ \bibinfo {pages} {054325} (\bibinfo
  {year} {2004})}\BibitemShut {NoStop}%
\bibitem [{\citenamefont {Navratil}\ and\ \citenamefont
  {Barrett}(1998)}]{Navratil98}%
  \BibitemOpen
  \bibfield  {author} {\bibinfo {author} {\bibfnamefont {P.}~\bibnamefont
  {Navratil}}\ and\ \bibinfo {author} {\bibfnamefont {B.~R.}\ \bibnamefont
  {Barrett}},\ }\href@noop {} {\bibfield  {journal} {\bibinfo  {journal} {Phys.
  Rev. C},\ }\textbf {\bibinfo {volume} {57}},\ \bibinfo {pages} {3119}
  (\bibinfo {year} {1998})}\BibitemShut {NoStop}%
\bibitem [{\citenamefont {Tilley}\ \emph {et~al.}(2004)\citenamefont {Tilley},
  \citenamefont {Kelley}, \citenamefont {Godwin}, \citenamefont {Millener},
  \citenamefont {Purcell}, \citenamefont {Sheu},\ and\ \citenamefont
  {Weller}}]{Tilley04}%
  \BibitemOpen
  \bibfield  {author} {\bibinfo {author} {\bibfnamefont {D.~R.}\ \bibnamefont
  {Tilley}}, \bibinfo {author} {\bibfnamefont {J.~H.}\ \bibnamefont {Kelley}},
  \bibinfo {author} {\bibfnamefont {J.~L.}\ \bibnamefont {Godwin}}, \bibinfo
  {author} {\bibfnamefont {D.~J.}\ \bibnamefont {Millener}}, \bibinfo {author}
  {\bibfnamefont {J.~E.}\ \bibnamefont {Purcell}}, \bibinfo {author}
  {\bibfnamefont {C.~G.}\ \bibnamefont {Sheu}}, \ and\ \bibinfo {author}
  {\bibfnamefont {H.~R.}\ \bibnamefont {Weller}},\ }\href@noop {} {\bibfield
  {journal} {\bibinfo  {journal} {Nucl. Phys. A},\ }\textbf {\bibinfo {volume}
  {745}},\ \bibinfo {pages} {155} (\bibinfo {year} {2004})}\BibitemShut
  {NoStop}%
\bibitem [{\citenamefont {Goldberg}\ \emph {et~al.}(1998)\citenamefont
  {Goldberg}, \citenamefont {Rogachev}, \citenamefont {Golovkov}, \citenamefont
  {Dukhanov}, \citenamefont {Serikov},\ and\ \citenamefont
  {Timofeev}}]{Goldberg98}%
  \BibitemOpen
  \bibfield  {author} {\bibinfo {author} {\bibfnamefont {V.~Z.}\ \bibnamefont
  {Goldberg}}, \bibinfo {author} {\bibfnamefont {G.~V.}\ \bibnamefont
  {Rogachev}}, \bibinfo {author} {\bibfnamefont {M.~S.}\ \bibnamefont
  {Golovkov}}, \bibinfo {author} {\bibfnamefont {V.~I.}\ \bibnamefont
  {Dukhanov}}, \bibinfo {author} {\bibfnamefont {I.~N.}\ \bibnamefont
  {Serikov}}, \ and\ \bibinfo {author} {\bibfnamefont {V.}~\bibnamefont
  {Timofeev}},\ }\href@noop {} {\bibfield  {journal} {\bibinfo  {journal} {JETP
  Lett.},\ }\textbf {\bibinfo {volume} {67}},\ \bibinfo {pages} {1013}
  (\bibinfo {year} {1998})}\BibitemShut {NoStop}%
\bibitem [{\citenamefont {Rogachev}\ \emph {et~al.}(2001)\citenamefont
  {Rogachev}, \citenamefont {Kolata}, \citenamefont {Becchetti} \emph
  {et~al.}}]{Rogachev01}%
  \BibitemOpen
  \bibfield  {author} {\bibinfo {author} {\bibfnamefont {G.~V.}\ \bibnamefont
  {Rogachev}}, \bibinfo {author} {\bibfnamefont {J.~J.}\ \bibnamefont
  {Kolata}}, \bibinfo {author} {\bibfnamefont {F.~D.}\ \bibnamefont
  {Becchetti}},  \emph {et~al.},\ }\href@noop {} {\bibfield  {journal}
  {\bibinfo  {journal} {Phys. Rev. C},\ }\textbf {\bibinfo {volume} {64}},\
  \bibinfo {pages} {061601(R)} (\bibinfo {year} {2001})}\BibitemShut {NoStop}%
\bibitem [{\citenamefont {Yamaguchi}\ \emph {et~al.}(2009)\citenamefont
  {Yamaguchi}, \citenamefont {Wakabayashi}, \citenamefont {Kubono} \emph
  {et~al.}}]{Yamaguchi09}%
  \BibitemOpen
  \bibfield  {author} {\bibinfo {author} {\bibfnamefont {H.}~\bibnamefont
  {Yamaguchi}}, \bibinfo {author} {\bibfnamefont {Y.}~\bibnamefont
  {Wakabayashi}}, \bibinfo {author} {\bibfnamefont {S.}~\bibnamefont {Kubono}},
   \emph {et~al.},\ }\href@noop {} {\bibfield  {journal} {\bibinfo  {journal}
  {Phys. Lett. B},\ }\textbf {\bibinfo {volume} {672}},\ \bibinfo {pages} {230}
  (\bibinfo {year} {2009})}\BibitemShut {NoStop}%
\bibitem [{\citenamefont {Halderson}(2004)}]{Halderson04}%
  \BibitemOpen
  \bibfield  {author} {\bibinfo {author} {\bibfnamefont {D.}~\bibnamefont
  {Halderson}},\ }\href@noop {} {\bibfield  {journal} {\bibinfo  {journal}
  {Phys. Rev. C},\ }\textbf {\bibinfo {volume} {69}},\ \bibinfo {pages}
  {014609} (\bibinfo {year} {2004})}\BibitemShut {NoStop}%
\bibitem [{\citenamefont {Volya}(2009)}]{Volya2009}%
  \BibitemOpen
  \bibfield  {author} {\bibinfo {author} {\bibfnamefont {A.}~\bibnamefont
  {Volya}},\ }\href@noop {} {\bibfield  {journal} {\bibinfo  {journal} {Phys.
  Rev. C},\ }\textbf {\bibinfo {volume} {79}},\ \bibinfo {pages} {044308}
  (\bibinfo {year} {2009})}\BibitemShut {NoStop}%
\bibitem [{\citenamefont {Rogachev}\ \emph {et~al.}(2010)\citenamefont
  {Rogachev}, \citenamefont {Johnson}, \citenamefont {Mitchell}, \citenamefont
  {Goldberg}, \citenamefont {Kemper},\ and\ \citenamefont
  {Wiedenh\"over}}]{Rogachev2010}%
  \BibitemOpen
  \bibfield  {author} {\bibinfo {author} {\bibfnamefont {G.~V.}\ \bibnamefont
  {Rogachev}}, \bibinfo {author} {\bibfnamefont {E.~D.}\ \bibnamefont
  {Johnson}}, \bibinfo {author} {\bibfnamefont {J.~P.}\ \bibnamefont
  {Mitchell}}, \bibinfo {author} {\bibfnamefont {V.~Z.}\ \bibnamefont
  {Goldberg}}, \bibinfo {author} {\bibfnamefont {K.~W.}\ \bibnamefont
  {Kemper}}, \ and\ \bibinfo {author} {\bibfnamefont {I.}~\bibnamefont
  {Wiedenh\"over}},\ }in\ \href@noop {} {\emph {\bibinfo {booktitle} {Fifth
  European Summer School on Experimental Nuclear Astrophysics}}},\ \bibinfo
  {series} {AIP Conference Proceedings}, Vol.\ \bibinfo {volume} {1213},\
  \bibinfo {editor} {edited by\ \bibinfo {editor} {\bibfnamefont
  {C.}~\bibnamefont {Spitaleri}}, \bibinfo {editor} {\bibfnamefont
  {C.}~\bibnamefont {Rolfs}}, \ and\ \bibinfo {editor} {\bibfnamefont {R.~G.}\
  \bibnamefont {Pizzone}}}\ (\bibinfo  {publisher} {American Institute of
  Physics},\ \bibinfo {year} {2010})\ pp.\ \bibinfo {pages}
  {137--148}\BibitemShut {NoStop}%
\bibitem [{\citenamefont {Warters}\ \emph {et~al.}(1953)\citenamefont
  {Warters}, \citenamefont {Fowler},\ and\ \citenamefont
  {Lauritsen}}]{Walters1953}%
  \BibitemOpen
  \bibfield  {author} {\bibinfo {author} {\bibfnamefont {W.~D.}\ \bibnamefont
  {Warters}}, \bibinfo {author} {\bibfnamefont {W.~A.}\ \bibnamefont {Fowler}},
  \ and\ \bibinfo {author} {\bibfnamefont {C.~C.}\ \bibnamefont {Lauritsen}},\
  }\href@noop {} {\bibfield  {journal} {\bibinfo  {journal} {Phys. Rev.},\
  }\textbf {\bibinfo {volume} {91}},\ \bibinfo {pages} {917} (\bibinfo {year}
  {1953})}\BibitemShut {NoStop}%
\bibitem [{\citenamefont {Greife}\ \emph {et~al.}(2007)\citenamefont {Greife}
  \emph {et~al.}}]{Greife2007}%
  \BibitemOpen
  \bibfield  {author} {\bibinfo {author} {\bibfnamefont {U.}~\bibnamefont
  {Greife}} \emph {et~al.},\ }\href@noop {} {\bibfield  {journal} {\bibinfo
  {journal} {Nucl. Inst. Meth. Phys. Res. B},\ }\textbf {\bibinfo {volume}
  {261}},\ \bibinfo {pages} {1089} (\bibinfo {year} {2007})}\BibitemShut
  {NoStop}%
\bibitem [{\citenamefont {Thompson}(1988)}]{fresco}%
  \BibitemOpen
  \bibfield  {author} {\bibinfo {author} {\bibfnamefont {I.~J.}\ \bibnamefont
  {Thompson}},\ }\href@noop {} {\bibfield  {journal} {\bibinfo  {journal}
  {Comp. Phys. Rep.},\ }\textbf {\bibinfo {volume} {7}},\ \bibinfo {pages}
  {167} (\bibinfo {year} {1988})}\BibitemShut {NoStop}%
\bibitem [{\citenamefont {Cohen}\ and\ \citenamefont
  {Kurath}(1965)}]{Cohen1965}%
  \BibitemOpen
  \bibfield  {author} {\bibinfo {author} {\bibfnamefont {S.}~\bibnamefont
  {Cohen}}\ and\ \bibinfo {author} {\bibfnamefont {D.}~\bibnamefont {Kurath}},\
  }\href@noop {} {\bibfield  {journal} {\bibinfo  {journal} {Nucl. Phys.},\
  }\textbf {\bibinfo {volume} {A73}},\ \bibinfo {pages} {1} (\bibinfo {year}
  {1965})}\BibitemShut {NoStop}%
\bibitem [{\citenamefont {Warburton}\ and\ \citenamefont
  {Brown}(1992)}]{Warburton1992}%
  \BibitemOpen
  \bibfield  {author} {\bibinfo {author} {\bibfnamefont {E.~K.}\ \bibnamefont
  {Warburton}}\ and\ \bibinfo {author} {\bibfnamefont {B.~A.}\ \bibnamefont
  {Brown}},\ }\href@noop {} {\bibfield  {journal} {\bibinfo  {journal} {Phys.
  Rev. C},\ }\textbf {\bibinfo {volume} {46}},\ \bibinfo {pages} {923}
  (\bibinfo {year} {1992})}\BibitemShut {NoStop}%
\bibitem [{\citenamefont {Navratil}\ \emph {et~al.}(2006)\citenamefont
  {Navratil}, \citenamefont {Bertulani},\ and\ \citenamefont
  {Caurier}}]{Navratil2006}%
  \BibitemOpen
  \bibfield  {author} {\bibinfo {author} {\bibfnamefont {P.}~\bibnamefont
  {Navratil}}, \bibinfo {author} {\bibfnamefont {C.~A.}\ \bibnamefont
  {Bertulani}}, \ and\ \bibinfo {author} {\bibfnamefont {E.}~\bibnamefont
  {Caurier}},\ }\href@noop {} {\bibfield  {journal} {\bibinfo  {journal} {Phys.
  Rev. C},\ }\textbf {\bibinfo {volume} {73}},\ \bibinfo {pages} {065801}
  (\bibinfo {year} {2006})}\BibitemShut {NoStop}%
\bibitem [{\citenamefont {Wiringa}\ \emph {et~al.}(2000)\citenamefont
  {Wiringa}, \citenamefont {Pieper}, \citenamefont {Carlson},\ and\
  \citenamefont {Pandharipande}}]{wiri00}%
  \BibitemOpen
  \bibfield  {author} {\bibinfo {author} {\bibfnamefont {R.~B.}\ \bibnamefont
  {Wiringa}}, \bibinfo {author} {\bibfnamefont {S.~C.}\ \bibnamefont {Pieper}},
  \bibinfo {author} {\bibfnamefont {J.}~\bibnamefont {Carlson}}, \ and\
  \bibinfo {author} {\bibfnamefont {V.~R.}\ \bibnamefont {Pandharipande}},\
  }\href@noop {} {\bibfield  {journal} {\bibinfo  {journal} {Phys.\ Rev.\ C},\
  }\textbf {\bibinfo {volume} {62}},\ \bibinfo {pages} {014001} (\bibinfo
  {year} {2000})}\BibitemShut {NoStop}%
\end{thebibliography}%

\end{document}